\documentclass[aps,prl,reprint]{revtex4-1}
\usepackage{amsmath,amssymb,graphicx}
\usepackage{amsbsy}
\usepackage{calrsfs}
\usepackage{latexsym}
\usepackage{color}
\usepackage{graphicx}
\usepackage{psfrag}
\usepackage[normalem]{ulem}
\usepackage{bm}
\usepackage{url}
\usepackage{notes2bib}
\usepackage{units}
\usepackage{soul}
\usepackage{dirtytalk}

\newcommand{\expv}[1]{ { \langle #1 \rangle } }

\begin{document}

%\title{Generosity as dominant strategy in multiplicative-growth environments: And God looked upon all that He made and, indeed, the growth rate was higher.}

\author{Lorenzo Fant$^{\dagger}$}
\email{lorenzofant@gmail.com}
\affiliation{International School for Advanced Studies (SISSA), Via Bonomea 265, 34136 Trieste, Italy}

\author{Onofrio Mazzarisi}
\altaffiliation{These authors contributed equally to this work}
\affiliation{Max Planck Institute for Mathematics in the Sciences, Inselstraße 22, 04103 Leipzig, Germany }

\author{Emanuele Panizon}
\altaffiliation{These authors contributed equally to this work}
\affiliation{Quantitative Life Sciences section, The Abdus Salam International Centre for Theoretical Physics (ICTP), Strada Costiera 11, 34014 Trieste, Italy}

\author{Jacopo Grilli}
\email{jgrilli@ictp.it}
\affiliation{Quantitative Life Sciences section, The Abdus Salam International Centre for Theoretical Physics (ICTP), Strada Costiera 11, 34014 Trieste, Italy}

\title{Stable cooperation emerges in stochastic multiplicative growth}

\begin{abstract}
Understanding the evolutionary stability of cooperation is a central problem in biology, sociology, and economics. There exist only a few known mechanisms that guarantee the existence of cooperation and its robustness to cheating.
Here, we introduce a new mechanism for the emergence of cooperation in the presence of fluctuations. We consider agents whose wealth changes
stochastically in a multiplicative fashion.
Each agent can share part of her wealth as public good, which is equally distributed among all the agents.
We show that, when agents operate with long time-horizons, cooperation produces an advantage at the individual level, as it effectively screens agents from the deleterious effect of environmental fluctuations. 
\end{abstract}

\maketitle

\textit{Introduction.---} The emergence and the stability of cooperation are a central problem in biology, sociology, and economics~\cite{Kropotkin1902,hamilton1964genetical,axelrod1981evolution,Bendor1997,Lehmann2006}.
Cooperation produces an advantage for the group, through the creation and sharing of social goods, but it is inherently unstable to cheating and to the tragedy of the commons,
where individual agents benefit from the social good without contributing to its creation~\cite{hardin1968tragedy,rankin2007tragedy}.
The dilemma of the evolution of cooperation can be solved in presence of some specific mechanisms~\cite{Nowak2006}, which lead to
the emergence and long-term stability of the cooperative trait. 

Many systems of interest for the study of cooperation exist in a context exposed to fluctuations and stochasticity.
A paradigmatic model for such systems, which has applications both in economics and in population biology, is geometric Brownian motion, which describes
the stochastic dynamics of a variable $x(t)$ as
$    \dot{x} = \mu x + \sigma x \xi(t) $,
where $\xi(t)$ is a delta-correlated white noise 
and $\mu$ and $\sigma$ are respectively drift and volatility.
In biology, $x$ could represent the abundance of a population, in economics $x$ is the value of an asset or the
wealth accumulated by a gambler. In the following we will refer to $x$ as wealth or value of an agent, keeping however in mind the breadth of possible applications of the geometric Brownian motion. 
An essential feature of multiplicative growth is that it lacks ergodicity, as the time-average behavior differs from the ensemble average~\cite{peters2013ergodicity}.
The latter grows exponentially in time with rate $\mu$, while the former grows with rate $g = \lim_{ t \to \infty} \langle \log x(t) \rangle / t = \mu-\sigma^2/2$.
This difference parallels the difference between arithmetic mean (which corresponds to the ensemble average) and geometric-mean (which converges to the time average), and it is the deep reason why the latter is a natural quantity to optimize for agents aiming at maximizing their future profits or growth.
In the context of gambling, the Kelly criterion defines the optimal size of a bet based on optimization of the geometric mean~\cite{kelly2011new}.
In evolutionary biology, under varying environmental conditions, natural selection favors traits on the basis of their geometric mean fitness~\cite{rivoire2011value,graves2017variability}.
An important consequence of the fact that the geometric mean fitness determines the optimal solution is that not only the average environment but also the amplitude of its fluctuations determine its value, as the geometric average grows with rate $\mu-\sigma^2/2$.
Reducing fluctuations, i.e., reducing the value of $\sigma$, has, therefore, a positive effect and should be expected to be advantaged by natural selection~\cite{melbinger2015impact}.

In the context of growth under fluctuating conditions, we introduce the possibility of asymmetric 
cooperation between $G$ agents with different drifts and volatilities, 
expanding on the settings of 
Refs.~\cite{bouchaud2000wealth,yaari2010cooperation,peters2015evolutionary,liebmann2017sharing,Stojkoski2019,Nowak2018,Amaral2021}.

Each agent shares - independently - part of her wealth as a public good, with a sharing rate $\alpha_i$, which is then divided equally among the agents. The presence of sharing couples the dynamics of agent's value. 

\begin{equation}
\dot{x}_i(t) = \mu_i x_i(t) + \sigma_i x_i(t) \xi_i(t) + \frac{1}{G} \sum_{j \neq i } \left( \alpha_j x_j(t)-\alpha_i x_i(t)\right) \ ,
\label{eq:DiffEq}
\end{equation}

where $\xi_i(t)$ are delta-correlated white noises, with an arbitrary correlation $\rho$ across agents. While we formulate the model as a diffusion process, its equivalency to time-discrete processes is discussed in detail in SM~\cite{Sup}  Sections~S1,2.

The full-defector scenario $\alpha_i = 0$ corresponds to the original Geometric Brownian motion solution $g_i = \mu_i - \sigma_i^2/2$.
If all the agents fully cooperate ($\alpha_i \to \infty$ for all $i$) and all the agents are equivalent ($\mu_i = \mu$ and $\sigma_i = \sigma$), one can obtain an exact solution of the trajectories $x_i(t)$~\cite{yaari2010cooperation,peters2015evolutionary,liebmann2017sharing,Stojkoski2019} resulting in an higher growth rate $g_i = \mu - \sigma^2/(2G)$. 
 The intuition behind these results is that, in this context, cooperation produces an advantage as it reduces effectively variability.
 By sharing their values with others, agents effectively diversify their investments, making their values less subject to fluctuations and, therefore, leading to faster growth.
However, this advantage alone does not explain how cooperation can emerge and why it could be stable to defection: also in the simple context of the prisoner dilemma, cooperation produces an individual advantage over defection, when all agents cooperate (i.e., cooperation is Pareto optimal).
The dilemma is, as well known, that cooperation is not stable (given that all the other agents are cooperating is advantageous for the individual to defect) while defection is (if all the agents are defecting there is no advantage in starting cooperating). 

In this Letter, we explore the stability and origin of cooperation under fluctuating conditions, using the setting of Eq.~(\ref{eq:DiffEq})
in the case of two agents ($G=2$).
We show that the maximization of the individual long-time return leads to the emergence and stability of cooperation.
We further explore the robustness of these results to uneven growth rates, correlated fluctuations, colored noise, costly cooperation, finiteness of time-horizons and extend them to arbitrary group sizes.
We show that, for large enough time-horizons, arbitrary levels of correlated fluctuations, noise time-correlation and costly sharing, cooperation (either full or partial) is advantageous at the individual level.
Finally, we explore the effect of these results on an explicit evolutionary dynamics.

\textit{Emergence and stability of cooperation.---} 
In order to make analytical progress on Eq.~(\ref{eq:DiffEq}) it is convenient to introduce $q_i(t) := \ln(x_i(t))$. The quantity that agents optimise is simply the typical growth rate $g_i = \lim_{t \to \infty} \langle q_i(t) \rangle/t$. The dynamics of $q_i$ can be obtained from Eq.~(\ref{eq:DiffEq}) using It\^{o} calculus (See SM~\cite{Sup}~S3).
One obtains
\begin{equation}
\expv{\dot{q}_1} = g_{0|0}^{(1)} - \frac{\alpha_1}{2} + \frac{\alpha_2}{2} \expv{\exp\left(q_2-q_1\right)}(t) \ ,
\label{eq:DiffEqLog}
\end{equation}
where $ g_{0|0}^{(i)} = \mu_i + \sigma_i^2/2$ is the growth rate in absence of cooperation.
The typical growth rate of agent $i$, in presence of another agent with resource sharing ratio $\alpha_j$, will therefore depend on both $\alpha_i$ and $\alpha_j$ and will be denoted by $g_{\alpha_i|\alpha_j}$. 
In the simple case of two agents, we can treat $g_{\alpha_i|\alpha_j}$ as the payoff function of a continuous game, aiming at finding the (pure-strategy) Nash equilibria and the evolutionary stable strategies.

We analitically derive (SM~\cite{Sup}~S3) that the dynamics of $\exp\left(q_2-q_1\right)$
 --- the only non trivial term in Eq.~(\ref{eq:DiffEqLog}) --- is ergodic with a stationary distribution, leading to a well defined term $\langle\exp\left(q_2-q_1\right)\rangle_{eq}$.
The growth rate $g^{(i)}_{\alpha_i|\alpha_j}$ will  be equal to $g^{(i)}_{0|0} + \left( \alpha_j \langle\exp\left(q_2-q_1\right)\rangle_{eq} - \alpha_i \right)/2 $. 
In particular, the effect of cooperation can be quantified by the difference ~\cite{bouchaud2015growth}
\begin{equation}
g^{(1)}_{\alpha_1|\alpha_2} - g^{(1)}_{0|0} = -\frac{\alpha_1}{2} + \frac{\sqrt{\alpha _1 \alpha _2}}{2} \frac{  K_{-1 + \gamma + \frac{\alpha _2-\alpha _1}{2 \sigma ^2}}\left(\frac{\sqrt{\alpha _1 \alpha _2}}{\sigma ^2}\right)}{ K_{-\gamma + \frac{\alpha _2-\alpha _1}{2 \sigma ^2}}\left(\frac{\sqrt{\alpha _1 \alpha _2}}{\sigma ^2}\right)} \ ,
\label{eq:gisolotauzero}
\end{equation}
where $K_{\beta}(z)$ is the modified Bessel function of the second type.
The value of $\sigma^2 \equiv (\sigma_{(1)}^2 + \sigma_{(2)}^2)(1-\rho)/2$ quantifies the effective magnitude of stochasticity (SM~\cite{Sup}~S4).
The parameter $\gamma$ is defined as the intrinsic difference of the uncoupled growth rates of the two agents with respect to stochasticity: $\gamma \equiv (g^{(1)}_{0|0} - g^{(2)}_{0|0} )/\sigma^2$. %We start by considering the simplest, yet non-trivial, case $\gamma = 0$.

% ADDED HERE
A relevant question now reads: given a strategy of the second player $\alpha_2$, what is the optimal value of $\alpha_1$? Mathematically, what is the value of $\alpha^{\ast} (\alpha_2) := \text{argmax}_{\alpha} g_{\alpha|\alpha_2}$ that maximizes the long-term growth rate as a function of the other agents' strategy? 

Morever, let us define an iterative process: agent 1 optimizes her sharing rate to maximize her own growth rate, given the sharing rate of the partner, following by agent 2, and again agent 1 and so on and so forth. What can we say about the \textit{evolutionary stable states} (ESSs) of their sharing rates $\alpha^{(i)}_{ess}$?

\begin{figure*}[t]
    \centering
    \includegraphics[width=1.\textwidth]{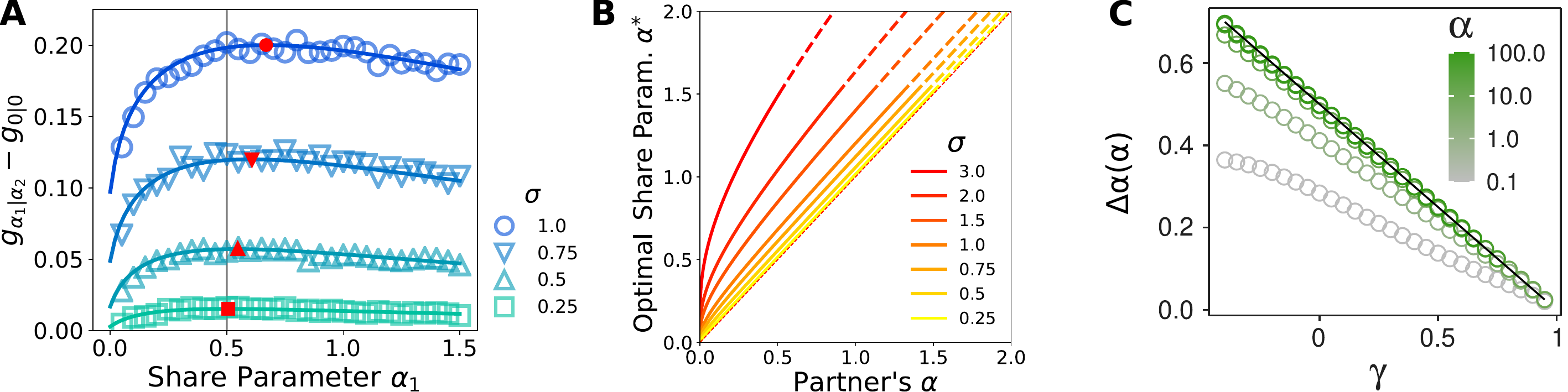}
    \caption{The optimal strategy for long term growth is to cooperate more than the partner as the growth rate of individual 1 is maximized by a value of $\alpha_1$ bigger than the partner's $\alpha_2$.
    Panel A displays the difference between the infinite time growth rate $g_{\alpha_1|\alpha_2}$  and the growth rate of the fully defecting case $g_{0|0}$ as a function of $\alpha_1$ for a fixed value $\alpha_2 = 0.5$ (vertical grey line). 
    Colors identify different noise amplitudes $\sigma$. 
    The analytical results of Eq.~(\ref{eq:gisolotauzero}) (solid lines) match the  numerical simulations (markers).  
    Red markers indicate the maxima of the curves, i.e. the value of $\alpha_1$ maximizing agent 1's growth rate. 
    Maxima appear on the right-hand side of the vertical grey line ($\alpha_2$), indicating that the optimal choice for agent 1 is to share more than agent 2.    
    Panel B shows the dependence on the partner's $\alpha$ of the optimal $\alpha^* := argmax(g_{\alpha_1|\alpha_2} - g_{0|0})$ that is the position of the maxima in panel A. Different lines represent different values of $\sigma$.
    Decreasing the noise amplitude the curves get closer to the diagonal while staying above it, showing that the optimal $\alpha$ is higher than the partner's $\alpha$ for any noise amplitude.
    In Panel C we represent the effect of intrinsic differences $\gamma$ on the optimal cooperation strategy. The plot shows the variation of cooperation rate $\Delta \alpha(\alpha)$ during a single step of the evolutionary process as a function of $\gamma$ and $\alpha$. 
    $\Delta\alpha > 0$ as long as $\gamma < 1$.
    That is, for $\gamma < 1$ both agents tend to increase their $\alpha$ in time for any value of $\alpha$.}
    \label{fig:fig1}
\end{figure*}

\textit{Absence of agents' intrinsic differences ($\gamma = 0$)---} 
%ADDED
We start by considering the simplest, yet non-trivial, case $\gamma = 0$.
%---
Fig.~\ref{fig:fig1} shows that our analytical solution of Eq.~(\ref{eq:gisolotauzero}) correctly matches the numerical simulations.
For fixed strategies $\alpha_1$ and $\alpha_2$, the effect of cooperation increases monotonically with $\sigma$: the higher are the fluctuations, the higher is the advantage of cooperation.
Interestingly, however, for a fixed value of $\alpha_2$, the long-term growth rate is not monotonic in $\alpha_1$.

The first non-trivial original result of our Letter is that the value of resource sharing that maximize the growth rate $\alpha_1^\ast(\alpha_2) %:= \text{argmax}_{\alpha_1} g_{\alpha_1|\alpha_2}
$ 
for a given strategy of the other agent $\alpha_2 >0 $ is always larger than the latter:
$\alpha_1^\ast(\alpha_2) > \alpha_2$ (Fig.~\ref{fig:fig1}).

Given the symmetry in the problem, the two agents converge - independently - to the same value $\alpha_{ess}$ of the sharing rate.
Particularly, for large $\alpha$, we find $\alpha^*(\alpha_2) \sim \alpha_2 + \sigma^2/4$, implying that $\alpha_{ess}\to \infty$.
This mathematical result implies that, contrarily to the mechanism in the tragedy of the commons,  each agent has an individual advantage in sharing \emph{more} than the other agent.
As a consequence, the evolutionary, adaptive, or learning dynamics maximizing the growth $g$ leads to a larger and larger level of cooperation (i.e., larger and larger values of $\alpha$).  

The intuition behind this result is that, in presence of fluctuations, sharing is akin to investment diversification, screening the agent from the detrimental effects of fluctuations.
In the long-time horizon, the return from this investment (the term $\alpha_2 \langle\exp\left(q_2-q_1\right)\rangle_{eq} / 2$) repays its cost (equal to $\alpha_1/2$).

%Complete defection remains an evolutionary fixed point however it is also unstable: As soon as both agents have any small sharing rate, they will converge to full cooperation. 
Complete defection remains a strict Nash equilibrium. However, as soon as both agents have any small sharing rate, they will escape from it and converge to full cooperation. In the definition of \cite{thomas1985evolutionarily}, it is not an ESS.

\textit{Presence of agents' intrinsic differences ($\gamma \neq 0$)---} 
The presence of $\gamma \neq 0$ in Eq.~(\ref{eq:gisolotauzero}) makes the agents' optimization non symmetric: under what conditions is it convenient for the agent with higher intrinsic growth rate to share with the other?

To investigate the evolution of $\alpha$ upon optimization we define $\Delta\alpha = \alpha^*_{\gamma}(\alpha^*_{-\gamma}(\alpha))$, that is the difference in the value of $\alpha$ on one individual after both individuals have asynchronously optimized their $\alpha$.
Figure~\ref{fig:fig1}C shows that if the intrinsic differences between the two agents are smaller than the level of noise ($|\gamma| < 1$) both the agents are expected to increase the sharing rate as $\Delta\alpha>0$ for every value of $\alpha$.
This implies that $\alpha_{ess}^{(i)} \to \infty$. 
However, as shown in SM~\cite{Sup}~S5, the presence of intrinsic differences between the two agents ($\gamma \neq 0$) corresponds to a not trivial ratio between the evolutionary stable share rates $\alpha_{ess}^{(1)}/\alpha_{ess}^{(2)} \to (1-\gamma)/(1+\gamma)$. 

On the contrary, when $|\gamma| > 1$ we find that the cooperative solution becomes evolutionary unstable and that evolution brings the value of $\alpha$ to decrease in time.
It thus becomes more advantageous for the better agent to defect, and the evolutionary stable strategy for both agents is $\alpha_{ess}^{(i)} = 0$.

\textit{Robustness of cooperation to neglected factors.---} 

Increasing the group size does not alter our results. 
Naively, increasing the group size $G$ could favor cheaters, that would take advantage of $G-1$ agents.
On the other hand, larger group sizes reduce even more the effect of fluctuations, providing a higher diversification.
In particular, the growth rate in the case of full cooperation converges to $\mu - \sigma^2/(2G)$~\cite{yaari2010cooperation,peters2015evolutionary,liebmann2017sharing,Stojkoski2019}.
As we numerically find (SM~\cite{Sup} Fig~S4), the latter effect is the dominates on the former, maintaining the evolutionary stability, while producing even higher advantages for cooperation.

Another key assumption is to describe fluctuations as white noise.
In reality we might expect, e.g. in biology~\cite{vasseur2004color} or economics~\cite{perello2002fat}, that fluctuations are time-correlated, over some timescale $\tau$, which could be comparable to the other timescales of the process.
We introduce this effect by assuming that fluctuations have an exponentially decaying auto-correlation
$\langle \xi_{i}(t) \xi_{i}(t') \rangle = \exp(-|t-t'|/\tau)/(2 \tau)$, reducing to white noise in the limit $\tau \to 0$.
While this case cannot be exactly solved, we approximated it using unified colored noise approximation~\cite{Jung1987}.
Our analytical approximation correctly matches numerical simulations for a wide range of values of $\tau$.
In particular, we obtain that full cooperation ($\alpha_{ess} \to \infty$) is not anymore an equilibrium strategy.
The optimal sharing rate turns out instead to depend on the value of $\tau$. For any value of $\tau$, the equilibrium sharing rate $\alpha_{ess}$ is a positive finite value.
For small values of $\tau$, the equilibrium sharing rate scales as
\begin{equation}
\alpha_{ess} \sim  \frac{\sigma\sqrt{1-\rho}}{2\sqrt{\tau}} \ ,
\end{equation}
which tends to full cooperation in the white-noise limit $\tau \to 0$.
Also, for a given value of $\tau$, larger levels of fluctuations and lower noise correlation produce increased cooperation.

In many settings cooperation is associated with a cost as seen, e.g., in microbes systems~\cite{rankin2007tragedy,Griffin2004,Damore2012,Cordero2012}. 
A cost can be introduced in multiple ways. 
If it is chosen to be proportional to the resource share rate, the model can be solved again analytically, showing that for any positive value of cooperation exists a finite equilibrium sharing rate (SM~\cite{Sup}~S5,S6).

A further key assumption of our framework was focusing on infinite-time horizons.
This can be relaxed by considering growth over finite time horizons $T$, and evaluating the average log-returns $\langle q_i(T) \rangle/T$.
This case is not amenable to analytical treatment 
but it can be studied in the context of evolutionary dynamics of finite populations discussed below.

\textit{Evolutionary dynamics in finite populations.---}The results presented above provide a clear mathematical mechanism for the emergence and stability of cooperation in the presence of fluctuations. 
In order to apply these results to a more concrete example, we perform an explicit evolutionary dynamics in a finite population.
We define a Wright-Fisher model with $N$ individuals (SM~\cite{Sup}~S8).
We associate the random variable $x_i(t)$ to the fitness and characterize individuals by their propensity to share $a_i\in[0,1]$,
the discrete counterpart to $\alpha_i$ (SM~\cite{Sup}~S7).
Individuals live in couples for a time $T$ and reproduce proportionally to the final fitness $x_i(T)$.

\begin{figure}[htbp]
    \centering
    \includegraphics[width=0.9\linewidth]{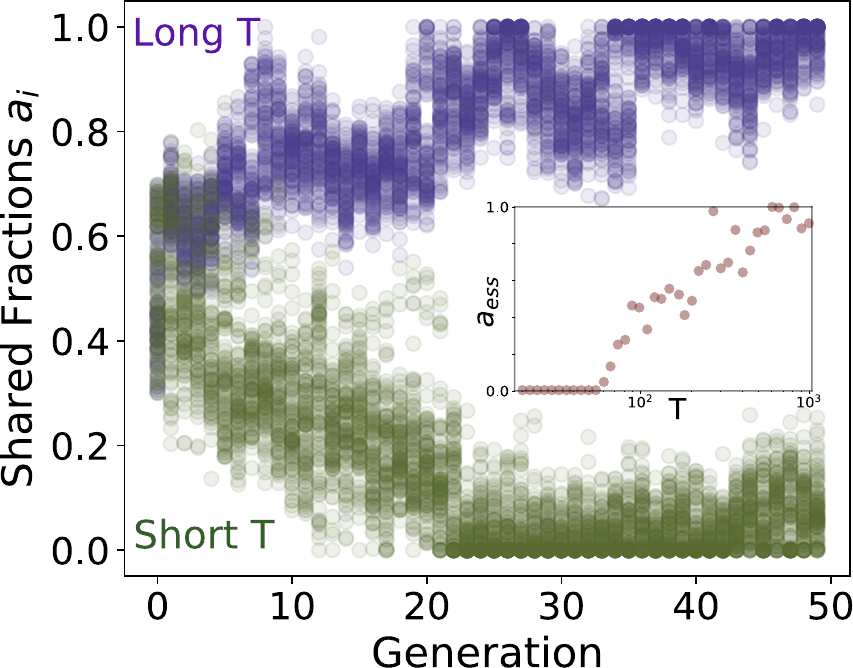}
    \caption{
    The emergence of cooperation in evolutionary dynamics depends on the time-horizon of the agents. 
    Two similar populations, with shared fractions $a_i$ initially distributed around $0.6$, evolve with different time horizons $T_{\textrm{short}}=20$ (green dots) and  $T_{\textrm{long}}=2000$ (blue dots).
    The population with a short time horizon evolves towards a distribution peaked in $a=0$ while the long time horizon one, in the opposite direction, towards a $a=1$.
    The inset shows the equilibrium value of the shared fraction $a_{ess}$, which displays a phase transition while increasing time horizon $T$ (measured in number of discrete time steps)}. 
    \label{fig:Evolution}
\end{figure}

As expected from previous results of population genetics in fluctuating environments~\cite{rivoire2011value}, evolution drives the population to traits that maximize the expected log-fitness. 
Figure~\ref{fig:Evolution} (main panel) shows the distribution of resource sharing probability $a$ over time.
For a short time horizon $T$, defection dominates and the distribution of $a$ is peaked close to $0$, with some variance due to mutations and genetic drift.
Conversely, for large enough time horizons, the vast majority of individuals cooperate, and $a$ peaks close to one.

The inset in Fig.~\ref{fig:Evolution} shows that the two regimes appear separated by a critical time horizon $T^*$ .
For $T > T^*$, the system behaves qualitatively as in the infinite time-horizon case: the individual optimizations of the log-average return lead agents to converge to a value $a_{ess} > 0$.
In particular, for very large time horizons we recover the prediction obtained under the diffusion limit and $a_{ess} \to 1$.
This result sheds light on the mechanism producing cooperation in our modeling setting: for long time-horizons, cooperation, thus investing in the other agents, continues providing returns, overcompensating its costs.

\textit{Discussion.---} In this Letter, we have discussed the optimal sharing strategy of agents in presence of multiplicative stochastic growth.
Cooperation can lead to faster growth of individual agents, therefore becoming an evolutionary stable strategy.
In this context, cooperation effectively screens agents from the detrimental effect of fluctuations: by cooperating, an agent effectively diversifies its investment, producing a higher return in the long term.
This is a sustainable strategy only if agents act with a long-time horizon and this altruistic investment have the time to repay off.
For short time-horizons, defection becomes again the evolutionary stable strategy.

Our approach differs considerably from previously identified mechanisms that explain the emergence and the stability of cooperation~\cite{Nowak2006}.
It does not in fact invoke multilevel selection (like in group or kin selection), as we consider individuals that only maximize their own growth rate.
Moreover, direct reciprocity~\cite{trivers1971evolution} is not the ingredient determining cooperation in our framework.
Direct reciprocity requires agents to change their actions based on the previous actions of other agents.
In our setting, given a value of the resource sharing rate $\alpha_2$ of agent 2, even if agent 1 is allowed to choose her own sharing rate $\alpha_1$ once for all, her optimal choice would be to share more than the other ($\alpha_1 > \alpha_2$).

The fundamental origin of the advantage of cooperation in our framework is due to the non-ergodicity of stochastic exponential growth, which effectively determines an individual advantage in reducing the level of fluctuations. 
Increasing the rate of cooperation comes at an immediate individual cost, as part of the wealth is diluted among agents as a public good, and has a long-term return, as the wealth shared with others is subject to independent fluctuations. 
The surprising result of this Letter is that the second effect is stronger than the first one, making it more advantageous --- at the individual level --- to cooperate.

It would be interesting to extend our framework in multiple directions.
We assume that individuals
share their value, but the scenario where they share only the income is potentially very interesting. 
We also assume that the group size is fixed but, for many biological (e.g., origin of multicellularity~\cite{szathmary1995major,bonner1998origins}) and sociological (e.g., group formation~\cite{reynolds1966open})
applications it would be interesting to treat it as a dynamical variable that can be optimised.
Finally we note that the results hold for unconstrained growth, which - while a reasonable assumption in the context of economic theory - may not apply in full generality to biological settings. Our results on finite time horizons (which effectively limit growth) suggest however that the main phenomenology could apply for different forms of constrained growth.

\begin{acknowledgments}
\section*{Acknowledgments}

We thank J.P. Bouchaud, M. Smerlak, M.A. Mu\~{n}oz and M. Cosentino Lagomarsino for insightful discussions and comments at various stages of the work and I. Macocco, M. Carli and T. Cossetto for the meetings that began our journey.
O.M. acknowledges the Alexander von Humboldt Foundation
in the framework of the Sofja Kovalevskaja Award
endowed by the German Federal Ministry of Education
and Research for providing funding for this work.
%We are grateful to Ettore Dore for inspiring ideas. 

\end{acknowledgments}
L. Fant, O. Mazzarisi and E. Panizon equally contributed to this work
\bibliography{bib}
\end{document}